\documentclass[a4paper,11pt]{article} 
\usepackage{jheppub}
\usepackage{amsmath} % AMS Math Package
\usepackage{amsthm}  % Theorem Formatting
\usepackage{amssymb} % Math symbols such as \mathbb 
\usepackage{slashed} 
\usepackage{hyperref}
\usepackage{pdfpages}
\usepackage{array}
\usepackage{soul}
\allowdisplaybreaks

\def\Mathematica{{{\sc Mathematica}}}
\def\Fiesta{{{\sc Fiesta 4.2}}}
\def\SecDec{{{\sc SecDec 3.0}}}
\def\FiniteFlow{{{\sc FiniteFlow}}}

\def\beq{\begin{equation}}
\def\eeq{\end{equation}}
\def\bea{\begin{eqnarray}}
\def\eea{\end{eqnarray}}
\def\beqn{\begin{eqnarray}} \def\eeqn{\end{eqnarray}}
\def\beeq{\begin{eqnarray}}
\def\eeeq{\end{eqnarray}}

\def\nn{\nonumber}
\def\Eq#1{Eq.~(\ref{#1})}

\pdfoutput=1

\newcommand{\valencia}{Instituto de F\'{\i}sica Corpuscular, Universitat de Val\`{e}ncia -- Consejo Superior de Investigaciones Cient\'{\i}ficas, Parc Cient\'{\i}fic, E-46980 Paterna, Valencia, Spain.}
\newcommand{\culiacan}{Facultad de Ciencias F\'isico-Matematicas, Universidad Aut\'onoma de Sinaloa, Ciudad Universitaria, CP 80000 Culiac\'an, Mexico.}

\begin{document}

\title{Causal representation of multi-loop Feynman integrands within the loop-tree duality}
\author[a]{J. Jes\'us Aguilera-Verdugo,}
\author[b]{Roger J. Hern\'andez-Pinto,}
\author[a]{Germ\'an Rodrigo,}
\author[a]{German~F. R.~Sborlini} 
\author[a]{and William~J.~Torres~Bobadilla}
\affiliation[a]{\valencia}
\affiliation[b]{\culiacan}

\emailAdd{jesus.aguilera@ific.uv.es}
\emailAdd{roger@uas.edu.mx}
\emailAdd{german.rodrigo@csic.es}
\emailAdd{german.sborlini@ific.uv.es}
\emailAdd{william.torres@ific.uv.es}

\preprint{IFIC/20-27}

\abstract{
The numerical evaluation of multi-loop scattering amplitudes in the Feynman representation 
usually requires to deal with both physical (causal) and unphysical (non-causal) singularities. 
The loop-tree duality (LTD) offers a powerful framework to easily characterise and 
distinguish these two types of singularities, and then simplify analytically the underling expressions.
In this paper, we work explicitly on the dual representation of multi-loop Feynman integrals
generated from three parent topologies, which we refer to as
Maximal, Next-to-Maximal and Next-to-Next-to-Maximal loop topologies. 
In particular, we aim at expressing these dual contributions, independently of the
number of loops and internal configurations, in terms of causal propagators only.
Thus, providing very compact and causal integrand representations to all orders. 
In order to do so, we reconstruct their analytic expressions from numerical evaluation over finite fields.
This procedure implicitly cancels out all unphysical singularities.
We also interpret the result in terms of entangled causal thresholds.
In view of the simple structure of the dual expressions, we integrate them numerically 
up to four loops in integer space-time dimensions, taking advantage of their smooth behaviour at integrand level.
}

\setcounter{page}{1}
\maketitle
%%%%%%%%%%%%%%%%%%%%%%%%%%%%%%%%%%%%%%%%%%%%%%%%%%%%%%%%%%%%%%
%
%				INTRODUCTION
%
%%%%%%%%%%%%%%%%%%%%%%%%%%%%%%%%%%%%%%%%%%%%%%%%%%%%%%%%%%%%%%

\section{Introduction}
\label{sec:introduction}

There is an important interest in considering higher order predictions  
in perturbation theory. This is due to the exceptional experimental measurements that 
have been done at the CERN's Large Hadron Collider (LHC)~\cite{Mangano:2020icy} and, also, in view of future colliders~\cite{
Abada:2019lih,Abada:2019zxq,Benedikt:2018csr,Abada:2019ono,
Blondel:2019vdq,Bambade:2019fyw,Roloff:2018dqu,CEPCStudyGroup:2018ghi}.  
In general, multi-loop integrals are not always well defined in $d = 4$ space-time dimensions. 
Hence, a special treatment needs to be carried out through dimensional 
regularisation (DREG)~\cite{Bollini:1972ui,tHooft:1972tcz}
that introduces extra difficulties.  
Several alternative techniques have appeared in the recent years~\cite{Gnendiger:2017pys}
that are aimed at overcoming this problem by keeping most of the computations in $d=4$ 
space-time dimensions.

Scattering amplitudes and loop integrals in the Feynman representation exhibit, in general, unphysical or non-causal singularities 
in the loop momentum space. Their representation in terms of Feynman parameters, Symanzik polynomials or Mellin-Barnes  
also inherit the unphysical structure of the original loop integral. However, all these unphysical singularities are expected to
cancel after integration.

It would be therefore desirable to work with loop representations, in which the integrand would exclusively exhibit the physical or causal singularities that characterise the final integrated expression. On the one hand, such representations would be more stable and faster to integrate through numerical methods. 
On the other hand, the presence of non-causal singularities obscures and makes difficult the analysis of the actual singular structure through, e.g., the Landau equations~\cite{Landau:1959fi}, particularly at higher quantum orders.

One of the most appealing features of the loop-tree duality (LTD)~\cite{Catani:2008xa,Bierenbaum:2010cy,Bierenbaum:2012th}
is that the unphysical or non-causal singularities cancel among different contributions of the integrand dual representation. 
This behaviour was firstly demonstrated for one-loop amplitudes~\cite{Buchta:2014dfa,Buchta:2015wna} 
and then to higher quantum orders~\cite{Aguilera-Verdugo:2019kbz},
and is the fundamental property that enables the simultaneous generation of loop and real-emission 
events through the four-dimensional unsubtraction (FDU)~\cite{Hernandez-Pinto:2015ysa,Sborlini:2016gbr,Sborlini:2016hat}. 
Other methods have also been proposed that are aimed at performing perturbative computations directly in the four 
physical space-time dimensions~\cite{Fazio:2014xea,Bobadilla:2016scr,BaetaScarpelli:2001ix,
BaetaScarpelli:2000zs,Pittau:2012zd,Page:2018ljf,Gnendiger:2017rfh,Bruque:2018bmy,Pozzorini:2020hkx}.

The LTD formalism has recently taken also a lot of attention from other authors~\cite{Tomboulis:2017rvd,Runkel:2019yrs,Runkel:2019zbm,Capatti:2019ypt}.
This is indeed due to the advantages of LTD in the numerical integration of loop integrals~\cite{Buchta:2015xda, Buchta:2015wna,Capatti:2019edf}
and, therefore, scattering amplitudes~\cite{Driencourt-Mangin:2017gop,Jurado:2017xut,Driencourt-Mangin:2019aix,Driencourt-Mangin:2019sfl,Driencourt-Mangin:2019yhu}.
In particular, an $N$-point amplitude at $L$ loops requires, within the LTD approach, only $\left(d-1\right)L$ integrations. 
Typically, this corresponds to the Euclidean space of the loop three-momenta, although LTD works as well in arbitrary coordinate systems. 
Since, in an Euclidean space, the hierarchies of scales are unambiguous, 
LTD has also been exploited as an alternative method~\cite{Driencourt-Mangin:2017gop,Plenter:2019jyj,Plenter:2020lop} 
for asymptotic expansions~\cite{Beneke:1997zp}.
Moreover, extra external momenta and, likewise, internal propagators do not alter the number of integrations.
Thus, the CPU time necessary for numerical integrations does not 
increase drastically with the number of external momenta as in other numerical approaches.

In a recent Letter~\cite{Verdugo:2020kzh}, 
we have conjectured that, in fact, LTD leads to loop integrand representations 
which are manifestly free of non-causal singularities to all orders and independently of the internal configuration.
In this paper, we elaborate further on the conjecture of Ref.~\cite{Verdugo:2020kzh}, which claims 
that non-causal singularities can be eliminated at integrand level from its analytic expression, then, 
leading to unintegrated representations displaying causal singularities exclusively. 
In order to achieve this conclusion, we rely on the classification of parent topologies 
at multi-loop level defined in Ref.~\cite{Verdugo:2020kzh}, which are used as the building block to describe more complex topologies.
In particular, two-loop scattering amplitudes can be characterised 
by a topology with 
$3$ 
momentum sets, where in each set there is an explicit dependence 
on the same loop momentum or a linear combination of the two loop momenta.
The generalisation of this topology at multi-loop level, 
with $L+1$ momentum sets,
is called Maximal Loop Topology~(MLT). 
Similarly, the extension to loop configurations with $L+2$ and $L+3$ sets, with $L$ arbitrary, originates 
the Next-to-Maximal Loop Topology~(NMLT) and the Next-to-Next-to-Maximal Loop Topology~(N$^2$MLT), respectively. 
Several relations exist among them through convolutions and factorisation identities, as presented in Ref.~\cite{Verdugo:2020kzh}.
These are the only topologies necessary to describe any scattering amplitude of up to three loops.
Beyond three loops, new topologies appear that are considered in another 
paper~\cite{Ramirez-Uribe:2020hes}.

Following this spirit, we present closed formulae for all the MLT, NMLT and N$^2$MLT multi-loop topologies
in terms of causal propagators only. We start from their compact LTD representations presented in Ref.~\cite{Verdugo:2020kzh}
that contain both causal and non-causal singularities.
Then, we reconstruct the full analytical result from numerical evaluations 
over finite fields~\cite{vonManteuffel:2014ixa,Peraro:2016wsq}.
This reconstruction algorithm allows us to overcome the non-causal 
propagators, since cancellations of the latter are implicitly performed.
To the best of our knowledge, the application of finite fields 
to generate integrands free of unphysical singularities is not present in
the literature and is studied here for the first time. On top of it, the algorithm 
presented in this paper can be straightforwardly extended to 
any topology, N$^k$MLT with $k>2$, and arbitrary internal configurations. 

For the purpose of elucidating the methods and results, we organise the paper in two parts.
The first one corresponds to the explicit study of the three parent loop topologies, 
MLT, NMLT and N$^2$MLT, first with single and then with multiple power propagators.
We emphasise on the physical causal structure that the integrands 
display as a byproduct of the LTD representation. 
Traditional approaches based on integration-by-parts identities~\cite{Chetyrkin:1981qh,Laporta:2001dd}
produce, in general, scalar integrals with powered propagators.
The LTD approach allows, without modifying our algorithm, to consider as well all these configurations, 
and therefore provides the interplay between our approach and the traditional methods.

The second part concerns to the study of the numerical performance of the causal 
LTD representation of the MLT, NMLT and N$^2$MLT topologies. 
We highlight that the structure of these compact formulae at $L$ loops
allows us to have a smooth and well-behaved numerical evaluation.
Certainly, with the presence of only physical singularities, we can elaborate more efficiently on the cancellation of the latter, 
following the lines of FDU~\cite{Hernandez-Pinto:2015ysa,Sborlini:2016gbr,Sborlini:2016hat}, 
in which the cancellation of infrared and ultraviolet singularities are performed at integrand level. 

The paper is structured as follows. In Sec.~\ref{sec:multiLTD}, we recall the main features 
of the all-order LTD representation with Lorentz-invariant infinitesimal 
complex prescription of dual propagators~\cite{Catani:2008xa,Bierenbaum:2010cy,Verdugo:2020kzh}.
Then, in Sec.~\ref{sec:analytic}, on top of the general expressions for the MLT, NMLT and N$^2$MLT 
configurations given in Ref.~\cite{Verdugo:2020kzh}, we present their compact analogous expressions 
in terms of exclusively causal propagators.  In order to obtain this set of expressions,
we use analytic reconstruction over finite fields through the
\verb"C++" implementation, \FiniteFlow~\cite{Peraro:2019svx}, of this algorithm
and interpret the results in terms of entangled causal thresholds.
In Sec.~\ref{sec:higher}, elaborating on Sec~\ref{sec:analytic}, we study topologies
with higher powers in the propagators.  We show that, independently of the powers 
of the propagators, we  end up with causal integrands. 
These results are independent of the number of space-time dimensions. 
Then, in Sec.~\ref{sec:numerical}, we numerically integrate the causal expressions obtained for
MLT, NMLT and N$^2$MLT with linear and raised powers in the propagators. 
In order to exploit the causal LTD representation, 
we perform several numerical tests at $d=2,3,4$ space-time dimensions, 
finding full agreement with softwares based on sector decomposition~\cite{Hepp:1966eg,Roth:1996pd,Binoth:2000ps,Heinrich:2008si}, 
\SecDec~\cite{Borowka:2015mxa} and \Fiesta~\cite{Smirnov:2015mct}.
Finally, in Sec.~\ref{sec:conclusions}, we draw our conclusions and future research directions. 

Algebraic manipulations of this paper have been 
carried out with an in-house \Mathematica~implementation of the LTD theorem.

%%%%%%%%%%%%%%%%%%%%%%%%%%%%%%%%%%%%%%%%%%%%%%%%%%%%%%%%%%%%%%
%
%				LTD
%
%%%%%%%%%%%%%%%%%%%%%%%%%%%%%%%%%%%%%%%%%%%%%%%%%%%%%%%%%%%%%%

\section{On the Loop-Tree Duality theorem to all orders}
\label{sec:multiLTD}

In this section, we set the notation and review the main features of the 
loop-tree duality (LTD) formalism. To start the discussion, 
let us consider a generic $N$-point scattering amplitude at $L$ loops,
\begin{equation}
\mathcal{A}_{N}^{\left(L\right)}\left(1,\hdots,n\right)=\int_{\ell_{1},\hdots,\ell_{L}}\sum\mathcal{N}\left(\left\{ \ell_{i}\right\} _{L},\left\{ p_{j}\right\} _{N}\right)\times G_{F}\left(1,\hdots,n\right)\,,
\label{eq:AmplitudGenerica}
\end{equation}  
where, to simplify the notation, we use the shorthand notation
 $\int_{\ell_{s}}\equiv-\imath\mu^{4-d}\int d^{d}\ell_{s}/\left(2\pi\right)^{d}$ 
for the integration measure,
where $\{\ell_i\}_{s=1,\ldots,L}$ corresponds to the independent loop momenta,
${\cal N}$ is a function of the loop and external momenta $\{p_j\}$, 
which is given by the Feynman rules of the theory. 
The function $G_F$ is defined as follows,
\begin{equation}
G_{F}\left(1,\ldots,n\right)=\prod_{i\in1\cup\cdots\cup n}\,(G_{F}(q_{i}))^{\alpha_{i}}\,,
\label{eq:ProductoGFs}
\end{equation}
with $\alpha_i$ the powers of the propagators and, 
\begin{align}
G_{F}\left(q_{i}\right) & =\frac{1}{q_{i}^{2}-m_{i}^{2}+\imath0}=\frac{1}{\left(q_{i,0}+q_{i,0}^{\left(+\right)}\right)\left(q_{i,0}-q_{i,0}^{\left(+\right)}\right)}\,,
\label{eq:gf}
\end{align}
the usual Feynman propagator of one single particle,
in which, $m_i$ corresponds to its mass, $\imath 0$ to the usual
infinitesimal imaginary prescription and,
\begin{align}
q_{i,0}^{\left(+\right)}=\sqrt{\mathbf{q}_{i}^{2}+m_{i}^{2}-\imath0}\,,
\end{align}
is the on-shell energy of the loop momentum $q_i$ 
written in terms of their spatial components $\textbf{q}_i$.
We have explicitly pulled out in~\Eq{eq:gf} the dependence of the Feynman propagator 
on the energy component of the loop momentum because we will integrate out this component explicitly.  
In general, the LTD theorem is defined in arbitrary coordinate systems,
as explained in Refs.~\cite{Catani:2008xa,Bierenbaum:2010cy,Verdugo:2020kzh}. 

The sum inside the integrand of Eq.~\eqref{eq:AmplitudGenerica} accounts 
for all the Feynman diagrams, or sets of Feynman diagrams, 
that contribute to the scattering amplitude. 
In the context of integration-by-parts identities~\cite{Chetyrkin:1981qh,Laporta:2001dd}, 
Eq~\eqref{eq:AmplitudGenerica} can represent a single Feynman integral
or a sum over master integrals. 

Since all the propagators are written in terms of independent loop momenta, 
we classify them through the flowing between them. 
With this in mind, the set $s$ includes all the propagators with
internal momenta of the form $q_{i_s}=\ell_s + k_{i_s}$, where $k_{i_s}$ 
is a linear combination of external momenta and $\ell_s$ is the loop momentum, 
or the linear combination of loop momenta, that characterises the set. 
The number of different sets $s$ is always larger than the number of loops beyond one loop. 
For the sake of simplicity, we will consider from now on that each set is composed by only one propagator.

In order to obtain the LTD representation for a given scattering amplitude, 
it is necessary to apply the Cauchy residue theorem iteratively and integrate out one 
degree of freedom for each loop momentum. 
Then, in each iterative step, we select the poles with negative imaginary part in the complex plane 
of the component of the loop momentum that is integrated.
In some intermediate steeps it is necessary to reverse some sets of momenta to keep a coherent momentum flow. 
A detailed discussion about this procedure is presented in Ref.~\cite{Aguilera-Verdugo:2020nrp}. 
Explicitly, starting from Eq.~(\ref{eq:AmplitudGenerica}) and setting on shell the propagators that 
depend on the first loop momentum, $q_{i_1}$, we define~\footnote{For each loop, and w.r.t. Ref.~\cite{Verdugo:2020kzh}, the factor $-2 \pi \imath$ 
from the Cauchy residue theorem is now included in the integration measure, \Eq{eq:measure}.}, 
\begin{equation}
{\cal A}_D^{(L)}(1;2,\ldots,n) \equiv \sum_{i_1 \in 1} {\rm Res}\left(d{\cal A}_N^{(L)}(1,\ldots,n),{\rm Im}(q_{i_1,0})<0 \right) \, ,
\label{eq:cauchy}
\end{equation} 
where $d{\cal A}_N^{(L)}$ is the integrand of the amplitude in \Eq{eq:AmplitudGenerica}.
The residue in \Eq{eq:cauchy} corresponds to integrate out the energy components of the loop momenta.
Hence, assuming that the iteration goes until the $r$-th set, we construct the nested residue as 
\begin{align}
{\cal A}_D(1,\ldots,r;r+1,\ldots,n) &\equiv
\sum_{i_r \in r} {\rm Res}\left({\cal A}_D(1,\ldots,r-1;r,\ldots,n),{\rm Im}(q_{i_{r},0})<0 \right) \,.
\label{eq:IteratedResidue}
\end{align}
The final LTD representation is given by the sum of all the nested residues and corresponds to setting simultaneously $L$ lines on shell,
which is equivalent to open the loop amplitude to non-disjoint trees.
In the following, we use the abbreviation
\begin{align}
\int_{\vec{\ell_{s}}}\,\bullet\equiv-\mu^{d-4}\int\frac{d^{d-1}\ell_{s}}{(2\pi)^{d-1}}\,\bullet\,,
\label{eq:measure}
\end{align}
for the $(d-1)$-momentum integration measure.

%%%%%%%%%%%%%%%%%%%%%%%%%%%%%%%%%%%%%%%%%%%%%%%%%%%%%%%%%%%%%%
%
%				FF RECONSTRUCTION
%
%%%%%%%%%%%%%%%%%%%%%%%%%%%%%%%%%%%%%%%%%%%%%%%%%%%%%%%%%%%%%%

\section{Causal representation of multi-loop integrals by analytic reconstruction}
\label{sec:analytic}

\begin{figure}[t]
\centering
\includegraphics[scale=0.9]{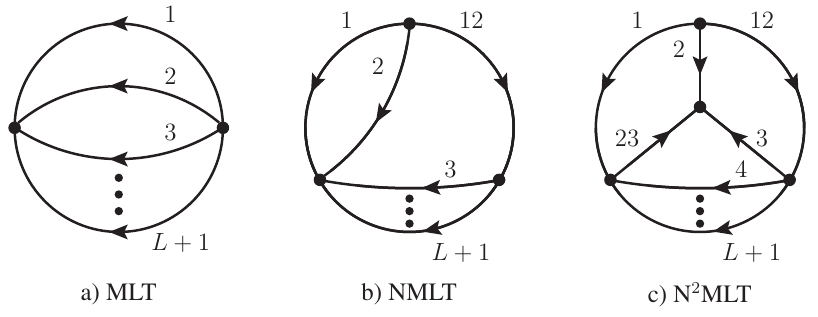}
\caption{Maximal Loop Topology (MLT), Next-to-Maximal Loop Topology (NMLT) and Next-to-Next-to-Maximal
Loop Topology (N$^2$MLT) at $L$ loops.
An arbitrary number of external momenta (not shown) is attached to the internal lines.}
\label{fig:alltopos}
\end{figure}

In former works~\cite{Aguilera-Verdugo:2019kbz}, we have demonstrated how 
the LTD formalism provides a comprehensive classification of causal singularities
and how unphysical ones are cancelled among paired terms.
These results stimulated the study of multi-loop topologies based on 
the definition of a systematic classification scheme, 
and leading to general LTD representations\footnote{A formal inductive proof of the all-order validity of such representations is presented in~\cite{Aguilera-Verdugo:2020nrp}.} 
of the MLT, NMLT and N$^2$MLT 
topologies depicted in Fig.~\ref{fig:alltopos}. 
In fact, along the same lines, 
these results have been extended to N$^4$MLT~\cite{Ramirez-Uribe:2020hes}.
In view of specific explicit examples, we also conjectured in Ref.~\cite{Aguilera-Verdugo:2019kbz} 
that LTD leads to integrand representations which are manifestly 
free of nonphysical singularities to all loop orders 
and regardless of the internal configuration.
Hence, here and in the successive sections, we elaborate on this 
conjecture and present causal representations of MLT, NMLT and N$^2$MLT to all loop orders. 

Starting from the compact LTD representations of the NMLT and N$^2$MLT multi-loop topologies
presented in Ref.~\cite{Verdugo:2020kzh} in terms of nested residues~\eqref{eq:IteratedResidue},  
that contain both causal and non-causal singularities, 
and motivated by their factorisation properties in terms of MLT subtopologies, for which we already obtained 
a causal representation, namely, free of non-causal singularities, 
we reconstruct in this section their full analytic expression in term of causal propagators only.
Here and in the following, we refer to causal propagators to denominators that only 
contain sums of on-shell energies of the loop momenta,
$q_{i,0}^{(+)}+q_{i+1,0}^{(+)}+q_{i+2,0}^{(+)}+\hdots+q_{m,0}^{(+)}$,
whereas non-causal ones are expressed as denominators 
that contain differences of
on-shell energies of the loop momenta,
$q_{i,0}^{(+)}-q_{i+1,0}^{(+)}+\hdots\pm q_{m,0}^{(+)}$,
leading thus to spurious singularities that are canceled as 
shall be described in this section. 
We perform this operation by numerical evaluation over finite fields~\cite{vonManteuffel:2014ixa,Peraro:2016wsq},
in which we use the \verb"C++" implementation of the \FiniteFlow~\cite{Peraro:2019svx} algorithm
together with its \Mathematica ~interface. 
In particular, we profit from the way how this algorithm solves linear systems.

Within the approach of reconstructing 
analytical expressions from numerical evaluations,
we simply end up with rational functions, whose variables are
the on-shell loop momenta  $q_{i,0}^{\left(+\right)}$ and the energy components of the external momenta, $p_{i,0}$. 
It turns out that this rational function is written only in terms of 
causal propagators~\cite{Verdugo:2020kzh}, which always have 
the structure of sums of on-shell loop energies.
This pattern, indeed, shows a very interesting behaviour, since 
the numerical evaluation of these quantities lacks of possible zeroes 
due to the absence of differences of $q_{i,0}^{\left(+\right)}$. 
As shall be described in the following, we comment more on this pattern,
elucidating how our final formula, originally constructed from non-causal 
propagators, contains only causal ones, and we interpret the result in terms 
of entangled causal thresholds.

Furthermore, it is important to obtain 
a closed formula that can describe the pattern of any topology with an  arbitrary number of loops. 
This, certainly, provides a parametric expression at all orders and, hence,
the calculation through the nested residue~\eqref{eq:IteratedResidue} is avoided. 
In order to sketch our procedure to obtain the causal analytic
expressions, we explicitly consider 
NMLT and  N$^{2}$MLT vacuum integrals first 
and then their generalisation with external momenta.

%%%%%%%%%%%%%%%%%%%%%%%%%%%%%%%%%%%%
\subsection{The two-loop sunrise diagram}

Before starting with the analysis of the multi-loop MLT, NMLT and N$^{2}$MLT configurations,
let us consider, for illustrative reasons, the simplest example of the two-loop sunrise diagram with three
propagators. Then, with the convention of~\eqref{eq:gf}, this two-loop integral becomes
\begin{align}
{\cal A}_2^{\left(2\right)} &  = \int_{\ell_{1}, \ell_{2}} G_F(1,2,12)
 =\int_{\ell_{1},\ell_{2}}\prod_{i=1,2,12}\frac{1}{\left(q_{i,0}-q_{i,0}^{\left(+\right)}\right)\left(q_{i,0}+q_{i,0}^{\left(+\right)}\right)}\,,
\label{eq:sun0}
\end{align}
with $q_{i}=\ell_{i}$ for $i\in\{1,2\}$,  and $q_{12}=-\ell_{1}-\ell_{2}+p$. Thus, by
applying the Cauchy residue theorem in $\left\{ \ell_{1},\ell_{2}\right\}$, we obtain the LTD representation of ${\cal A}_{2}^{\left(2\right)}$ 
in terms of the nested residues~\eqref{eq:IteratedResidue}, 
\begin{align}
{\cal A}_2^{\left(2\right)}=\int_{\vec{\ell}_{1},\vec{\ell}_{2}}\left[G_{D}\left(1,2\right)+G_{D}\left(1,\overline{12}\right)+G_{D}\left(\overline{2},\overline{12}\right)\right]\,,
\label{eq:sun}
\end{align} 
where the bar indicates a reversal of the momentum flow 
of the corresponding propagators, $q_{\overline i} = -q_i$, 
and $G_D$ represents the double residue of the integrand of ${\cal A}_2^{(2)}$,
\begin{align}
G_{D}\left(i, j\right)\equiv \text{Res} \left( \text{Res}
\left( G_F(1,2,12) ,\left\{q_{i,0} = q_{i,0}^{\left(+\right)}\right\} \right),\left\{ q_{j,0} = q_{j,0}^{\left(+\right)}\right\} \right)\,.
\end{align}
Then, the first term of the integrand in Eq.~\eqref{eq:sun},
\begin{align}
G_{D}\left(1,2\right)= & \frac{1}{4\, q_{1,0}^{\left(+\right)}q_{2,0}^{\left(+\right)}
\left(q_{1,0}^{\left(+\right)}+q_{2,0}^{\left(+\right)} - p_0+q_{12,0}^{\left(+\right)}\right)
\left(q_{1,0}^{\left(+\right)}+q_{2,0}^{\left(+\right)} - p_0 -q_{12,0}^{\left(+\right)}\right)}\,,
\end{align} 
allows us to implement the decomposition
\begin{align}
G_{D}\left(1,2\right)= & - \frac{1}{8\, q_{1,0}^{\left(+\right)}q_{2,0}^{\left(+\right)}q_{12,0}^{\left(+\right)}}\left(
\frac{1}{q_{1,0}^{\left(+\right)}+q_{2,0}^{\left(+\right)}+q_{12,0}^{\left(+\right)}-p_0}-
\frac{1}{q_{1,0}^{\left(+\right)}+q_{2,0}^{\left(+\right)}-q_{12,0}^{\left(+\right)}-p_0}\right)\,,
\label{eq:decomp}
\end{align}
where the first term on the r.h.s. of \Eq{eq:decomp} represents the double residue over the positive energy mode of 
the propagator with momentum $q_{12}$.  This term generates a causal threshold at $\sum_{i=1,2,12} q_{i,0}^{(+)} = p_0$, if $p_0>0$, 
when $q_{12}$ becomes on shell, and represents a configuration where the three on-shell momenta are aligned in the same direction.
The second term in \Eq{eq:decomp} is non-causal and shall cancel. Hence, by applying partial fractioning to the three contributions of~\eqref{eq:sun}, 
we notice that non-causal denominators are piecewise canceled, leading to,
\begin{align}
{\cal A}_{2}^{\left(2\right)} = - \int_{\vec{\ell}_{1},\vec{\ell}_{2}}
\frac{1}{8q_{1,0}^{\left(+\right)}q_{2,0}^{\left(+\right)}q_{12,0}^{\left(+\right)}} \left( 
\frac{1}{q_{1,0}^{\left(+\right)}+q_{2,0}^{\left(+\right)}+q_{12,0}^{\left(+\right)}-p_0} +
\frac{1}{q_{1,0}^{\left(+\right)}+q_{2,0}^{\left(+\right)}+q_{12,0}^{\left(+\right)}+p_0}
\right) \,,
\label{eq:sun2}
\end{align}
which is manifestly free of non-causal thresholds. Besides the causal threshold which is active if $p_0>0$, this expression 
also generates the complementary causal threshold for $p_0<0$  that arise when the three on-shell momenta flow in 
the opposite direction. In other words, when the three negative energy modes are selected and set on shell. If $p_0=0$, both threshold 
configurations occur simultaneously, although only the limit $q_{i,0}^{(+)} \to 0$, with $i\in\{1,2,12\}$, can lead to an integrand 
singularity. 

We would like to remark that the ordering in which the residues are taken
does not alter the final result given by \Eq{eq:sun2}, although individual terms are
modified. For example, the expression 
\begin{align}
{\cal A}_{2}^{\left(2\right)}&=\int_{\vec{\ell}_{1},\vec{\ell}_{2}}
\left[G_D\left(1,12\right)+G_D\left(1,\overline 2\right)+G_D\left(\overline 2,\overline{12}\right)\right]\,,
\end{align}
leads exactly to \Eq{eq:sun2}, although the individual contributions
are different from those in \Eq{eq:sun} and also exhibits non-causal thresholds. 

%%%%%%%%%%%%%%%%%%%%%%%%%%%%%%%%%%%%%%%%%%%%%%%
\begin{figure}[htb!]
\centering
\includegraphics[scale=0.59]{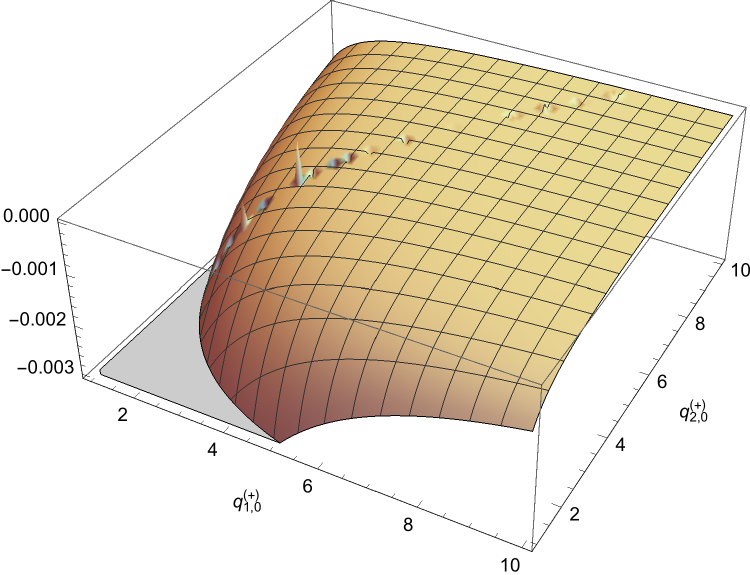}
\includegraphics[scale=0.59]{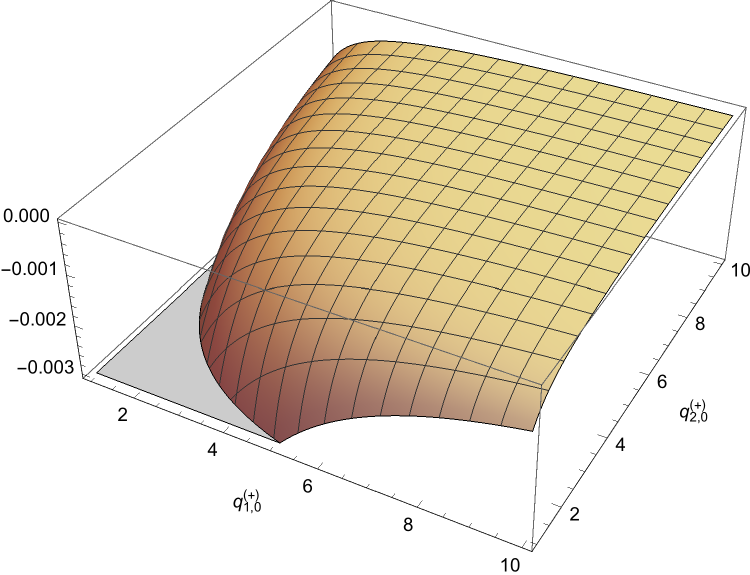}
\caption{Three dimensional plots for the integrand of the two-loop sunrise diagram
in terms of non-causal (left) and causal (right) propagators. The numerical fluctuations due to numerical cancellations 
of non-causal thresholds are visible on the left plot. The right plot is stable because the integrand expression 
is manifestly free of non-causal thresholds.}
\label{fig:dplots}
\end{figure}
%%%%%%%%%%%%%%%%%%%%%%%%%%%%%%%%%%%%%%%%%%%

Let us now briefly comment on the structure of the two kind of integrands.
Both representations, with or without non-causal thresholds, are physically equivalent.
However, the LTD representation that still contains non-causal denominators is not optimal
since it exhibits spurious singularities. 
This is illustrated in Fig.~\ref{fig:dplots}, where we show the difference 
between the full expression~\eqref{eq:sun} and the simplified one~\eqref{eq:sun2}.
The small peaks in the left plot are due to numerical fluctuations introduced by 
the numerical cancellation of non-causal thresholds
arising at $q_{1,0}^{\left(+\right)}\pm q_{2,0}^{\left(+\right)}\mp q_{12,0}^{\left(+\right)}=0$ (assuming $p_0 = 0$),
which correspond to the singularities of the individual terms in~\eqref{eq:sun}. 
 
%%%%%%%%%%%%%%%%%%%%%%%%%%%%%%%%%%%%%%%%%%
\subsection{Maximal Loop Topology}
\label{nonvacuumMLT}

The MLT configuration is characterised by $L+1$ sets of propagators with an arbitrary number of
propagators in each set. A pictorial representation of this topology is given in Fig.~\ref{fig:alltopos}\textcolor{blue}{a}.
Its general LTD representation is given by~\cite{Verdugo:2020kzh},
\begin{equation} 
{\cal A}_{\text{MLT}}^{\left(L\right)}\left(1,2,\hdots,L+1 \right) = \int_{\vec{\ell}_{1},\cdots,\vec{\ell}_{L}}
\sum_{i=1}^{L+1} {\cal A}_D(1, \ldots, i-1, \overline{i+1}, \ldots, \overline{L+1}; i)\,,
\end{equation}
 in terms of the nested residues defined in \Eq{eq:sun}.
This expression is valid for any loop integral or scattering amplitude with an arbitrary internal configuration,
although it contains both causal and non-causal thresholds. 
Again, the bar in $\overline i$ indicates a reversal of the momentum flow or equivalently a selection 
of negative energy modes.

For the sake of simplicity, we will consider in the following only one propagator in each set 
and scalar integrals.
The momenta of these propagators are defined as follows.
\begin{align}
q_{i}=\ell_{i}\,,\qquad\text{with }i \in \{1,\hdots,L\}\,,&&q_{L+1}=-\sum_{i=1}^{L}\ell_{i} - p_1\,.
\label{eq:setMLT}
\end{align}
The two-loop integral ${\cal A}_{2}^{\left(2\right)}$ in Eq.~\eqref{eq:sun0} indeed belongs to the MLT family.
Likewise, any two-loop amplitude can be cast in this way, independently on whether 
it contains planar or non-planar topologies. At this order, $L=2$, there are only three independent sets of propagators. 

%%%%%%%%%%%%%%%%%%%%%%%%%%%%%%%%%%%%%%%%%%
\begin{figure}[t]
\centering
\includegraphics[scale=0.9]{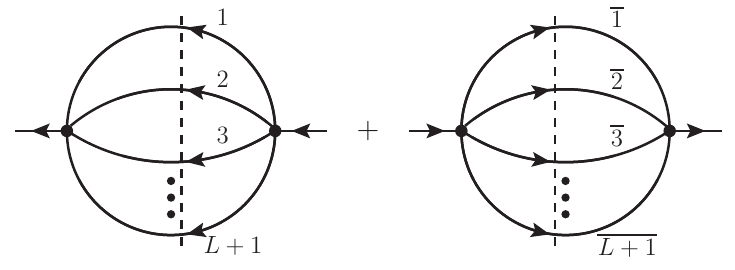}
\caption{Causal thresholds of the MLT topology. The arrow on the external momentum represents the positive energy flow.}
\label{fig:mlt}
\end{figure}
%%%%%%%%%%%%%%%%%%%%%%%%%%%%%%%%%%%%%%%%%%

A causal representation of this topology was already presented in Ref.~\cite{Verdugo:2020kzh},
\begin{align}
{\cal A}_{\text{MLT}}^{\left(L\right)}\left(1,2,\hdots,(L+1)_{-p_{1}}\right)=
&-\int_{\vec{\ell}_{1},\cdots,\vec{\ell}_{L}}\frac{1}{x_{L+1}}\left(\frac{1}{\lambda_1^{-}}+\frac{1}{\lambda_1^{+}}\right)\,,
\label{eq:fullmltp}
\end{align}
where $(L+1)_{-p_1}$ means that the external momentum $p_1$ has been inserted in the loop momentum $L+1$
according to~\eqref{eq:setMLT}, $x_{L+k}=2^{L+k}\prod_{i=1}^{L+k}q_{i,0}^{\left(+\right)}$ and,
\begin{align}
\lambda_1^\pm=\sum_{i=1}^{L+1}q_{i,0}^{\left(+\right)} \pm p_{1,0}\,.
\end{align}
Due to the simplicity that this set of integrals holds, we have used it as a first test of the reconstruction algorithm that 
is explained later. By setting $L=2$, we recover~\eqref{eq:sun2}.

Each of the two terms of the integrand in \Eq{eq:fullmltp} represents a potential causal threshold singularity.
Only one of them is active once the sign of the energy of the external momentum, $p_{1,0}$, is fixed. 
These two causal thresholds can be interpreted exactly as we interpreted \Eq{eq:sun2}. 
The threshold singularities arise when all the momenta are on shell and either aligned in one direction or the opposite one. 
The graphical interpretation of these causal thresholds is illustrated in Fig.~\ref{fig:mlt}.

%%%%%%%%%%%%%%%%%%%%%%%%%%%%%%%%%%%%%%%%%%
\subsection{NMLT vacuum integral}
\label{vacuumNMLT}

As we have observed, the MLT topology is sufficient to describe any two-loop configuration. Starting from three loops, 
the NMLT and N$^2$MLT topologies are also necessary to characterise the loop configurations that are not 
described by MLT. General LTD representations for NMLT and N$^2$MLT have been presented 
in Ref.~\cite{Verdugo:2020kzh} that contain both causal and non-causal thresholds. 
In this section, we consider their causal LTD representation. To simplify the presentation, we start by considering 
configurations with one single propagator in each set, and no external momenta. The more complex case with external momenta
will be considered in Sec.~\ref{withexternal}.
Then, to describe NMLT configurations, on top of considering the internal momenta~\eqref{eq:setMLT}, with $p_1=0$,
we also need to add an additional one, 
\begin{align}
q_{L+2}=-\ell_{1}-\ell_{2} \,.\label{eq:setnmlt}
\end{align}
A pictorial representation of NMLT is provided in Fig~\ref{fig:alltopos}\textcolor{blue}{b}.

The causal LTD representation that we obtain is
\begin{align}
{\cal A}_{\text{NMLT}}^{\left(L\right)}\left(1,2,\hdots,L+2\right)=&\int_{\vec{\ell}_{1},\hdots,\vec{\ell}_{L}}\frac{2}{x_{L+2}}\frac{\sum_{i=1}^{L+2}q_{i,0}^{\left(+\right)}}{\lambda_{1}\lambda_{2}\lambda_{3}}\,.
\label{eq:fullnmlt0}
\end{align}
with 
\begin{align}
\lambda_{1}=\sum_{i=1}^{L+1}q_{i,0}^{\left(+\right)}\, & & \lambda_{2}=q_{1,0}^{\left(+\right)}+q_{2,0}^{\left(+\right)}+q_{L+2,0}^{\left(+\right)}\,, 
& & \lambda_{3}=\sum_{i=3}^{L+2}q_{i,0}^{\left(+\right)}\,.
\label{eq:l1tol3}
\end{align}
This expression, although it is slightly more complicated than the MLT one,
can easily be rewritten by partial fractioning in the more suitable form,
\begin{align}
{\cal A}_{\text{NMLT}}^{\left(L\right)}\left(1,2,\hdots,L+2\right)=&\int_{\vec{\ell}_{1},\hdots,\vec{\ell}_{L}}\frac{2}{x_{L+2}}\left(\frac{1}{\lambda_{1}\lambda_{2}}+\frac{1}{\lambda_{2}\lambda_{3}}+\frac{1}{\lambda_{3}\lambda_{1}}\right)\,.
\label{eq:fullnmlt}
\end{align}

The analytic expression of Eq.~\eqref{eq:fullnmlt} can alternatively be reconstructed from 
numerical evaluations over finite fields. Since the integrand in~\eqref{eq:fullnmlt0} is a rational function
in the on-shell energies of the loop momenta, $q_{i,0}^{(+)}$,
we find relations among $q_{i,0}^{\left(+\right)}$ and $\lambda_i$.
For instance, at $L=3$, we find
\begin{align}
 & q_{1,0}^{\left(+\right)}=\frac{1}{2}\left(\lambda_{1}+\lambda_{2}-\lambda_{3}-2q_{2,0}^{\left(+\right)}\right)\,, &  & q_{5,0}^{\left(+\right)}=\frac{1}{2}\left(-\lambda_{1}+\lambda_{2}+\lambda_{3}\right)\,.\nonumber \\
 & q_{3,0}^{\left(+\right)}=\frac{1}{2}\left(\lambda_{1}-\lambda_{2}+\lambda_{3}-2q_{4,0}^{\left(+\right)}\right)\,,
\end{align}
Then, by plugging these relations in~\eqref{eq:fullnmlt0}, we directly find~\eqref{eq:fullnmlt}. 
Because the numerator of~\eqref{eq:fullnmlt0} is a linear function of $q_{i,0}^{\left(+\right)}$,
both approaches are of the same difficulty. 

In this particular case, it is straightforward to observe the linear relations between 
$\lambda_i$ and $q_{i,0}^{(+)}$, leading, in this way, to simplifications originated by
polynomial divisions. In fact, as stated above, Eq.~\eqref{eq:fullnmlt0} can be directly obtained
by using partial fractioning. In more general cases, partial fractioning 
will produce more terms that need to be properly canceled. 
Hence, to avoid the proliferation of terms at intermediate steps, we make use of the 
analytic reconstruction over finite fields to obtain compact expressions
containing only causal propagators. 
In the following section, we shall note that simplifications are not straightforward 
because the degree of the numerator we want to reduce increases as the number
of causal propagators. 

%%%%%%%%%%%%%%%%%%%%%%%%%%%%%%%%%%%%%%%%%%
\begin{figure}[t]
\centering
\includegraphics[scale=0.9]{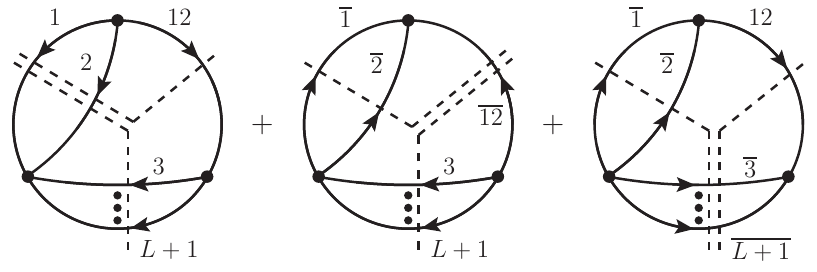}
\caption{Entangled causal thresholds of the NMLT topology. External momenta not shown.}
\label{fig:nmlt}
\end{figure}
%%%%%%%%%%%%%%%%%%%%%%%%%%%%%%%%%%%%%%%%%%

Let us now interpret \eqref{eq:fullnmlt} in terms of what we call \textit{entangled causal thresholds}.
Each of the $\lambda_i$ represents a potential causal threshold singularity that, 
as in the MLT case, requires that all the momentum flows are aligned in the same direction. 
The product of two causal denominators can be understood as representing physical configurations 
where two sets of propagators can simultaneously go on shell. For this to happen, the common propagators 
have to be in the same configuration. This entanglement can also be understood from the factorisation 
identity that NMLT fulfils in terms of MLT subtopologies, as explained in Ref.~\cite{Verdugo:2020kzh}. 

A pictorial interpretation of the entangled causal structure of~\eqref{eq:fullnmlt} is provided in Fig.~\ref{fig:nmlt}, 
where the dashed lines single out the internal propagators that eventually go on shell simultaneously;
a subset of them is already on shell through LTD. 
Each diagram in Fig.~\ref{fig:nmlt} has two dashed lines that
correspond to the two denominators $\lambda_i$ and $\lambda_j$ which
are present in each term of~\eqref{eq:fullnmlt}. The causal thresholds are entangled because the 
momentum flow of the propagators that are common to both causal thresholds are matched. 
For example, the first diagram of Fig.~\ref{fig:nmlt} represents the term $1/(\lambda_1 \lambda_2)$
corresponding to the causal thresholds $\{1,2,\hdots,L\}$ and $\{1,2,L+1\}$ that share the 
entangled propagators $\{1,2\}$.

%%%%%%%%%%%%%%%%%%%%%%%%%%%%%%%%%%%%%%%%%%
\subsection{N$^{2}$MLT vacuum integral}
\label{vacuumN2MLT}

The N$^2$MLT is the last topology that is needed to describe any full scattering amplitude at three loops. 
In fact, N$^2$MLT is the master topology that describes also MLT and NMLT 
configurations to all orders~\cite{Verdugo:2020kzh}. 
This configuration is depicted in Fig.~\ref{fig:alltopos}\textcolor{blue}{c}, and  is usually called 
Mercedes--Benz topology. 
Similar to the NMLT case, besides considering the internal momenta~\eqref{eq:setMLT}
and~\eqref{eq:setnmlt}, we add one more, 
\begin{align}
q_{L+3}=-\ell_{2}-\ell_{3} \,.\label{eq:setn2mlt}
\end{align}
For the moment, we consider configurations without external momenta.
Then, with $L+3$ internal propagators, we have all the required ingredients to 
understand the structure of any integral at three loops and, equivalently, any scattering amplitude.

Hence, from the LTD representation of N$^{2}$MLT given in Ref.~\cite{Verdugo:2020kzh}, 
we end up with the integrand written as a rational function, 
\begin{align}
{\cal A}_{\text{N}^2\text{MLT}}^{\left(L\right)}\left(1,2,\hdots,L+3\right) &= \int_{\vec{\ell}_{1},\hdots,\vec{\ell}_{L}}
\frac{1}{x_{L+3}}\frac{N(\{q_{i,0}^{(+)}\})}{\prod_{i=1}^{7}\lambda_{i}}\,,
\end{align}
with $\lambda_1$ through $\lambda_3$ defined in \Eq{eq:l1tol3}, 
\begin{align}
 & \lambda_{4}=q_{2,0}^{\left(+\right)}+q_{3,0}^{\left(+\right)}+q_{L+3,0}^{\left(+\right)}\,, &  & \lambda_{6}=q_{1,0}^{\left(+\right)}+q_{3,0}^{\left(+\right)}+q_{L+2,0}^{\left(+\right)}+q_{L+3,0}^{\left(+\right)}\,,\nonumber \\
 & \lambda_{5}=q_{1,0}^{\left(+\right)}+q_{L+3,0}^{\left(+\right)}+\sum_{i=4}^{L+1}q_{i,0}^{\left(+\right)}\,, &  & \lambda_{7}=q_{2,0}^{\left(+\right)}+\sum_{i=4}^{L+3}q_{i,0}^{\left(+\right)}\,,
\end{align}
and $N(\{ q_{i,0}^{(+)}\})$ a degree-four polynomial in $q_{i,0}^{(+)}$.
From the structure of the latter it is not straightforward to make
manifest an expression with the features of Eqs.~(\ref{eq:fullmltp}) and (\ref{eq:fullnmlt}). 
Thus, we reconstruct its analytic expression through finite fields.
In fact, one notices that the denominators $\lambda_{i}$
are not independent; they, instead, obey a few relations. Hence, we write some $q_{i,0}^{\left(+\right)}$
and $\lambda_{1}$ in terms of $\lambda_{i}$, 
\begin{align}
 & q_{1,0}^{\left(+\right)}=\frac{1}{2}\left(\lambda_{2}+\lambda_{5}-\lambda_{7}\right)\,, &  & q_{6,0}^{\left(+\right)}=\frac{1}{2}\left(-\lambda_{4}-\lambda_{5}+\lambda_{6}+\lambda_{7}\right)\,,\nonumber \\
 & q_{2,0}^{\left(+\right)}=\frac{1}{2}\left(\lambda_{2}+\lambda_{4}-\lambda_{6}\right)\,, &  & q_{7,0}^{\left(+\right)}=\frac{1}{2}\left(-\lambda_{2}-\lambda_{3}+\lambda_{6}+\lambda_{7}\right)\,,\nonumber \\
 & q_{3,0}^{\left(+\right)}=\frac{1}{2}\left(\lambda_{3}+\lambda_{4}-\lambda_{7}\right)\,, &  & \lambda_{1}=\lambda_{2}+\lambda_{3}+\lambda_{4}+\lambda_{5}-\lambda_{6}-\lambda_{7}\,.\nonumber \\
 & q_{4,0}^{\left(+\right)}=\frac{1}{2}\left(\lambda_{3}+\lambda_{5}-\lambda_{6}-2q_{5,0}^{\left(+\right)}\right)\,,
\end{align}
Then, by properly replacing  $q_{i,0}^{\left(+\right)}$ and $\lambda_{i}$,
according to their relations
and performing a straightforward polynomial division, we find, 
\begin{align}
{\cal A}_{\text{N}^{2}\text{MLT}}^{\left(L\right)}\left(1,2,\hdots,L+3\right)=&-\int_{\vec{\ell}_{1},\cdots,\vec{\ell}_{L}}\frac{2}{x_{L+3}}\Bigg[\frac{1}{\lambda_{1}}\left(\frac{1}{\lambda_{2}}+\frac{1}{\lambda_{3}}\right)\left(\frac{1}{\lambda_{4}}+\frac{1}{\lambda_{5}}\right)\nonumber\\
&+\frac{1}{\lambda_{6}}\left(\frac{1}{\lambda_{2}}+\frac{1}{\lambda_{4}}\right)\left(\frac{1}{\lambda_{3}}+\frac{1}{\lambda_{5}}\right)+\frac{1}{\lambda_{7}}\left(\frac{1}{\lambda_{2}}+\frac{1}{\lambda_{5}}\right)\left(\frac{1}{\lambda_{3}}+\frac{1}{\lambda_{4}}\right)\Bigg]\,.
\label{eq:fulln2mlt}
\end{align}

%%%%%%%%%%%%%%%%%%%%%%%%%%%%%%%%%%%%%%%%%%
\begin{figure}[t]
\centering
\includegraphics[scale=0.9]{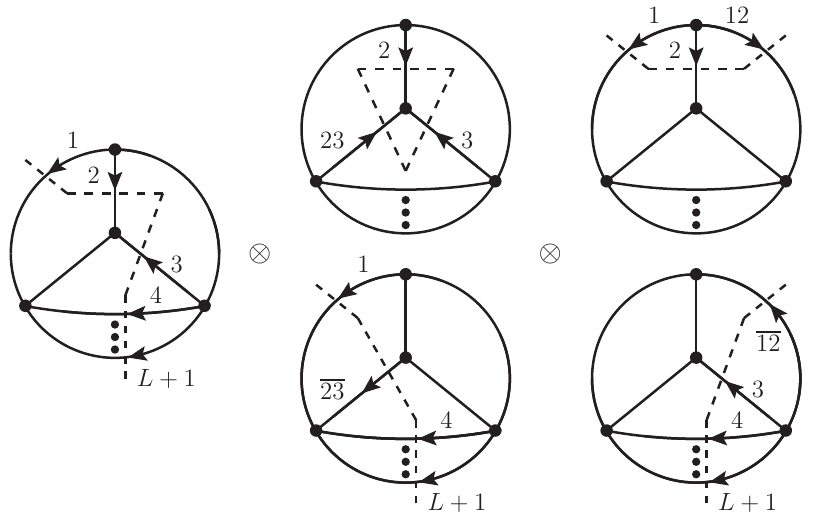}
\caption{Entangled causal thresholds of the N$^2$MLT topology. External momenta not shown.}
\label{fig:n2mlt}
\end{figure}
%%%%%%%%%%%%%%%%%%%%%%%%%%%%%%%%%%%%%%%%%%

Analogous to the NMLT case, we note that~\eqref{eq:fulln2mlt}
is written to all orders in term of causal propagators only, in this case in terms of the 
product of three causal propagators.  This pattern 
can be understood from two approaches. First, from their factorisation 
properties in terms of MLT and NMLT configurations, as explained in Ref.~\cite{Verdugo:2020kzh}.
Second, from the entanglement of three causal thresholds. 
A pictorial representation of this entanglement is presented in Fig.~\ref{fig:n2mlt}
for the first term in the r.h.s. of~\eqref{eq:fulln2mlt}.

%%%%%%%%%%%%%%%%%%%%%%%%%%%%%%%%%%%%%%%%%%
\subsection{NMLT and N$^{2}$MLT topologies with external momenta}
\label{withexternal}

In this section, we show how to generalise the causal representations 
of the NMLT and N$^2$MLT vacuum diagrams presented in 
Sections~\ref{vacuumNMLT} and \ref{vacuumN2MLT}
to the most general case that considers insertion of external momenta.
Then, to obtain analytic and compact expressions for these two
topologies, we follow the same algorithm based on finite fields. 
The vacuum expression in  Eqs.~\eqref{eq:fullnmlt} and~\eqref{eq:fulln2mlt} have been used to guide this computation.
Let us anticipate that the insertion of external momenta does not affect
the causal physical behaviour of these integrals. The only difference now is that, 
as for the causal MLT representation given in Section~\ref{nonvacuumMLT},  we have to 
distinguish the entangled configurations that correspond to external momenta with 
positive or negative energy flow, or in other words, if the external momenta are incoming or outgoing. 

\begin{figure}[t]
\centering
\includegraphics[scale=0.7]{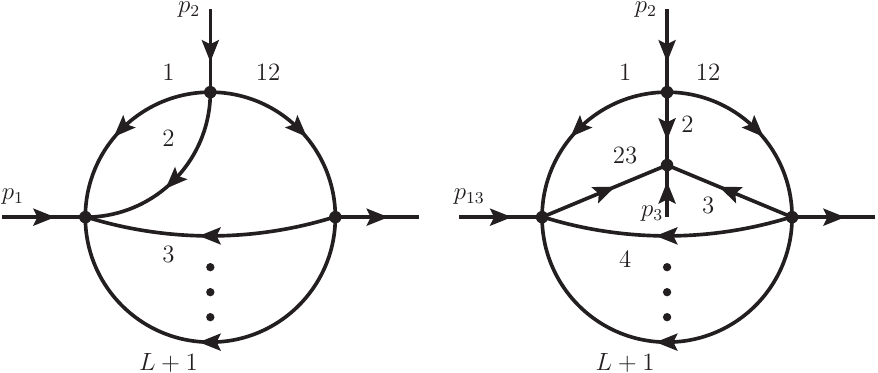}
\caption{Next-to-Maximal Loop Topology (left) and Next-to-Next-to-Maximal
Loop Topology (right) at $L$ loops 
with the insertion of external momenta $p_i$,
according to Eqs.~\eqref{eq:extmom}. 
}
\label{fig:alltoposp}
\end{figure}

We can apply the same procedure for the insertion of the external  momenta $p_1$, $p_2$ and $p_3$ 
in the internal momenta $L+1$, $L+2$ and $L+3$, respectively, 
\begin{equation}
 q_{L+1}=-\sum_{i=1}^{L+1} \ell_{i} - p_{13}\,, \qquad
 q_{L+2}=-\ell_{1}-\ell_{2} + p_2\,, \qquad
 q_{L+3}=-\ell_{2}-\ell_{3} - p_3\,. 
 \label{eq:extmom}
\end{equation}
The three external momenta $p_i$ are considered to have positive energy when they are incoming.
By momentum conservation, we should also have either $p_{12} = p_1+p_2$ for NMLT
or $p_{123} = \sum_{i=1}^3 p_i$ for N$^2$MLT as outgoing momentum in one of the vertices. 
Besides, we emphasise that we set $p_3=0$ for NMLT while keeping the definition 
of the internal momenta at~\eqref{eq:extmom}.
These topologies, with the insertion of external momenta, are depicted in Fig.~\ref{fig:alltoposp}.

We note that the causal propagators $\lambda_i$ are now shifted by the external momenta $\pm p_i$ or a linear 
combination of them as 
\begin{align}
& \lambda_{1}^\pm =\sum_{i=1}^{L+1} q_{i,0}^{\left(+\right)} \pm p_{13,0}\,,  && \nn \\
& \lambda_{2}^\pm =q_{1,0}^{\left(+\right)}+q_{2,0}^{\left(+\right)}+q_{L+2,0}^{\left(+\right)} \pm p_{2,0}\,, 
& & \lambda_{3}^\pm=\sum_{i=3}^{L+2}q_{i,0}^{\left(+\right)} \mp p_{123,0} \,, \nn \\
& \lambda_{4}^\pm=q_{2,0}^{\left(+\right)}+q_{3,0}^{\left(+\right)}+q_{L+3,0}^{\left(+\right)} \pm p_{3,0}\,, 
& & \lambda_{6}^\pm =q_{1,0}^{\left(+\right)}+q_{3,0}^{\left(+\right)}+q_{L+2,0}^{\left(+\right)}+q_{L+3,0}^{\left(+\right)} \pm p_{23,0}\,,\nonumber \\
& \lambda_{5}^\pm=q_{1,0}^{\left(+\right)}+q_{L+3,0}^{\left(+\right)}+\sum_{i=4}^{L+1}q_{i,0}^{\left(+\right)} \pm p_{1,0}\,, 
& & \lambda_{7}^\pm=q_{2,0}^{\left(+\right)}+\sum_{i=4}^{L+3}q_{i,0}^{\left(+\right)} \pm p_{12,0}\,.
\end{align}
With the formula that we present in the following, we can describe, with a single representation, 
up to three-point functions for NMLT and up to four-point functions for N$^2$MLT.

The causal representation of NMLT is a function of $\lambda_1^\pm$ through $\lambda_3^\pm$, with $p_3=0$, and is given by 
\begin{align}
 {\cal A}_{\text{NMLT}}^{\left(L\right)}(1,2,\hdots,(L+1)_{-p_{1}},&(L+2)_{p_2})=\int_{\vec{\ell}_{1},\cdots,\vec{\ell}_{L}}\frac{1}{x_{L+2}} \nn \\
&  \times \bigg[\frac{1}{\lambda_{1}^{+} \lambda_{2}^{-}}+\frac{1}{\lambda_{2}^{+} \lambda_{3}^{-}}
+ \frac{1}{\lambda_{3}^{+} \lambda_{1}^{-}} + (\lambda_i^+\leftrightarrow \lambda_i^-) \bigg]\,.
\end{align}
Due to the insertion of external momenta, we have now to consider the entangled threshold configurations that distinguish if the external 
momenta are incoming or outgoing, namely, if their energy flow is positive or negative. With our conventions, positive energy flows
correspond to incoming momenta.  The exchange $\lambda_i^+\leftrightarrow \lambda_i^-$ accounts for the configurations 
with opposite momentum flows and results in a doubling of the terms obtained for the vacuum diagrams. 

The causal N$^2$MLT representation also exhibits a very compact expression 
\begin{align}
{\cal A}_{\text{N}^{2}\text{MLT}}^{\left(L\right)}\big(1,2 & ,\hdots,\left(L+1\right)_{-p_{13}},\left(L+2\right)_{p_{2}},\left(L+3\right)_{-p_{3}}\big)=-\int_{\vec{\ell}_{1},\cdots,\vec{\ell}_{L}}\frac{1}{x_{L+3}}\nonumber \\
\times & \Bigg[\frac{1}{\lambda_{1}^{+}}\left(\frac{1}{\lambda_{2}^{-}}+\frac{1}{\lambda_{3}^{-}}\right)\left(\frac{1}{\lambda_{4}^{+}}+\frac{1}{\lambda_{5}^{+}}\right)+\frac{1}{\lambda_{6}^{+}}\left(\frac{1}{\lambda_{3}^{-}}+\frac{1}{\lambda_{5}^{-}}\right)\left(\frac{1}{\lambda_{2}^{+}}+\frac{1}{\lambda_{4}^{+}}\right)\nonumber \\
 & +\frac{1}{\lambda_{7}^{+}}\left(\frac{1}{\lambda_{3}^{-}}+\frac{1}{\lambda_{4}^{-}}\right)\left(\frac{1}{\lambda_{2}^{+}}+\frac{1}{\lambda_{5}^{+}}\right)+\left(\lambda_{i}^{+}\leftrightarrow\lambda_{i}^{-}\right)\Bigg]\,.
\end{align}
 
Let us briefly summarise on the algorithm used to compute all these formulae. 
As mentioned before, we profit from the software 
\FiniteFlow~and its built-in functions, \verb"FFLinearFit" and \verb"FFDenseSolve",
to analytically reconstruct the rational function of the configuration and 
to find relations between $q_{i,0}^{\left(+\right)}$ and $\lambda_{i}$, respectively. 
We would like to remark that these expressions for MLT, NMLT and N$^{2}$MLT
have been analytically checked, with the iterated application of the LTD
theorem, up to six loops, finding completely agreement. 
The pattern displayed by all topologies allows us to generalise 
and provide a closed formula that has the mathematical support
of the studies carried out in~\cite{Verdugo:2020kzh,Aguilera-Verdugo:2020nrp} to all orders.
In other words, only causal contributions remain in the final expressions, being all
the non-causal or unphysical terms cancelled at intermediate steps. 
We also note that the structure of the integrands suggests a smooth numerical
evaluation, which we profit in Sec.~\ref{sec:numerical}. Although we have presented explicit 
expressions only for scalar integrals, the algorithm is valid as well for non-scalar integrals.

%%%%%%%%%%%%%%%%%%%%%%%%%%%%%%%%%%%%%%%%%%%%%%%%%%%%%%%%%%%%%%
%
%				RAISED PROPAGATORS
%
%%%%%%%%%%%%%%%%%%%%%%%%%%%%%%%%%%%%%%%%%%%%%%%%%%%%%%%%%%%%%%

\section{Topologies with higher powers in the propagators}
\label{sec:higher}

So far we have discussed the structure of MLT, NMLT and N$^{2}$MLT configurations in which 
the dependence on the Feynman propagators is linear. However,
in practical applications, like UV local renormalisation~\cite{Driencourt-Mangin:2017gop,Driencourt-Mangin:2019aix,Driencourt-Mangin:2019yhu} or multi-loop
calculations, one also deals with higher powers in the propagators~\cite{Chetyrkin:1981qh,Laporta:2001dd,Bierenbaum:2012th}.
In this section, we elaborate on the compact formulae, found in the previous section,
and provide a procedure for computing $L$-loop integrals with higher powers in the propagators. 
In particular, due to the causal structure observed in all the cases,
it is expected that this physical behaviour should also remain with higher powers of the propagators. 
Abusing of the notation and recalling we are considering only one propagator in each loop set, let us define,
\begin{align}
{\cal A}_{\rm N^{k-1}MLT}^{\left(L\right)}\left(1^{\alpha_{1}},2^{\alpha_{2}},\hdots,(L+k)^{\alpha_{L+k}}\right)&=\int_{\ell_{1},\ldots,\ell_{L}}\mathcal{N}\times\prod_{i=1}^{L+k}\left(G_{F}\left(q_{i}\right)\right)^{\alpha_{i}}\,,
\label{eq:powers}
\end{align}
with $k \in \{1,2,3\}$, for any $L$-loop integral, where the superscript $\alpha_i $ corresponds 
to the power of the $i$-th propagator. In the following, for the sake of the simplicity, we restrict our study 
to the scalar case in which the numerator is $\mathcal{N}=1$. 

Remarkably, due to the structure of the Feynman propagator (\ref{eq:gf}), it is straightforward to raise
the power in the propagators by simple performing $(\alpha_i-1)$ derivatives w.r.t. $q_{i,0}^{\left(+\right)}$, 
\begin{align}
\left( G_{F}\left(q_{i}\right) \right)^{\alpha_i}=\frac{1}{\left(\alpha_i-1\right)!}\frac{\partial^{\alpha_i-1}}{\partial\left((q_{i,0}^{\left(+\right)})^{2}\right)^{\alpha_i-1}}
\, G_{F} (q_i)\,.
\end{align}
Therefore, the results obtained in Sec.~\ref{sec:analytic} can be used for the
purpose of the present discussion. 
Furthermore, we stress that the expressions obtained in the present paper
are valid in any dimension, since only the energy components of the loop momenta 
have been integrated. Hence, we can yet
numerically evaluate these integrals in any integer dimensions.
Since the LTD representation of the N$^{k-1}$MLT configurations that we have considered are manifestly free of non-causal propagators,
the corresponding loop integrals with raised propagators will also be causal. 
In the following we will consider integrals that are ultraviolet and infrared finite
in order to check the better numerical performance of the causal LTD representation. 

Therefore, if we were to evaluate
finite integrals, for instance in integer space-time dimensions, 
it is sufficient to consider the causal LTD representation with linear propagators
and perform the derivatives in the on-shell energies $q_{i,0}^{\left(+\right)}$.
For example, the MLT configuration, 
with the insertion of an external momentum, 
\begin{align}
{\cal A}_{\text{MLT}}^{\left(L\right)}\left(1^{2},2^{2},\hdots,L^{2},(L+1)_{-p_1}\right)&=\prod_{i=1}^{L}\frac{\partial}{\partial(q_{i,0}^{\left(+\right)})^{2}} {\cal A}_{\text{MLT}}^{\left(L\right)}\left(1,2,\hdots,(L+1)_{-p_1}\right)\,.
\end{align}

In order to elucidate the operation of raising powers, 
we consider the simplest case 
${\cal A}_{\text{MLT}}^{\left(L\right)}\left(1^{2},2,\hdots,(L+1)_{-p_1}\right)$ 
with
one single squared propagator. Since the denominators depend linearly 
on $q_{i,0}^{\left(+\right)}$, we utilise the chain rule as follows, 
\begin{align}
\frac{\partial}{\partial(q_{i,0}^{\left(+\right)})^{2}}\,\bullet & =\frac{1}{2q_{i,0}^{\left(+\right)}}\frac{\partial}{\partial(q_{i,0}^{\left(+\right)})}\,\bullet\,.
\end{align}
This amounts to,
\begin{align}
{\cal A}_{\text{MLT}}^{\left(L\right)}\left(1^{2},2,\hdots,\left(L+1\right)_{-p_{1}}\right) & =\frac{1}{2q_{1,0}^{\left(+\right)}}\frac{\partial}{\partial(q_{1,0}^{\left(+\right)})}{\cal A}_{\text{MLT}}^{\left(L\right)}\left(1^{2},2,\hdots,\left(L+1\right)_{-p_{1}}\right)\,,\nonumber \\
 & =\int_{\vec{\ell}_{1},\hdots,\vec{\ell}_{L}}\frac{1}{2q_{1,0}^{\left(+\right)}x_{L+1}\,}\left[\frac{1}{q_{1,0}^{\left(+\right)}}\left(\frac{1}{\lambda_{1}^{+}}+\frac{1}{\lambda_{1}^{-}}\right)+\frac{1}{(\lambda_{1}^{+})^{2}}+\frac{1}{(\lambda_{1}^{-})^{2}}\right]
\end{align}

In the following, we present the numerical evaluation of several MLT, NMLT and N$^{2}$MLT configurations.
We remark that, within LTD, the inclusion of internal masses do not stem any issue.
In more details, one needs to perform $(d-1)$ integrations for each loop. 
The number of integrations turn out to be lower than approaches based on Feynman parametrisation.

%%%%%%%%%%%%%%%%%%%%%%%%%%%%%%%%%%%%%%%%%%%%%%%%%%%%%%%%%%%%%%
%
%				INTEGRATION
%
%%%%%%%%%%%%%%%%%%%%%%%%%%%%%%%%%%%%%%%%%%%%%%%%%%%%%%%%%%%%%%

\section{Numerical evaluation of N$^{k-1}$MLT}
\label{sec:numerical}

In view of the results of Secs.~\ref{sec:analytic} and \ref{sec:higher}, we elaborate 
on those expressions by numerically integrating in the $(d-1)$-loop momenta, $\mathbf{q}_{i}$.
This is indeed done to investigate the stability of this set
of formulae, written now in terms of causal propagators only.
To this end, we evaluate multi-loop integrals in $d=2,3,4$
space-time dimensions~\footnote{The case with $d=1$ space-time dimensions~\cite{Runkel:2019zbm} is trivial.
One just replaces $q_{i,0}^{\left(+\right)}\to\sqrt{m_{i}^{2}}$. },
presenting results for topologies with and without higher powers in the propagators
up to four loops. 

The numerical results presented in this section are double checked with the softwares \SecDec~and 
\Fiesta.

%%%%%%%%%%%%%%%%%%%%%%%%%%%%%%%%%%%%%
\subsection{Two-dimensional integrals}
\label{sec:2d-int}

\begin{figure}[thb!]
\centering
\includegraphics[scale=0.8]{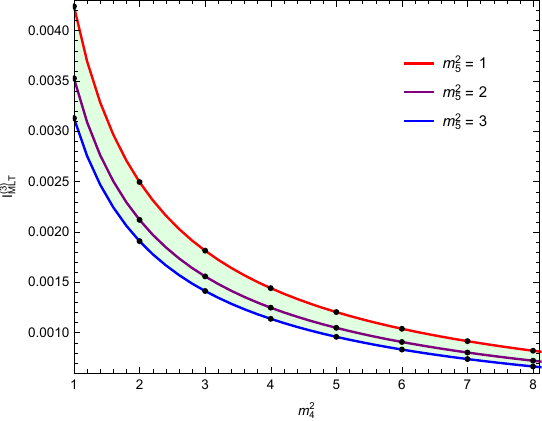}
\includegraphics[scale=0.8]{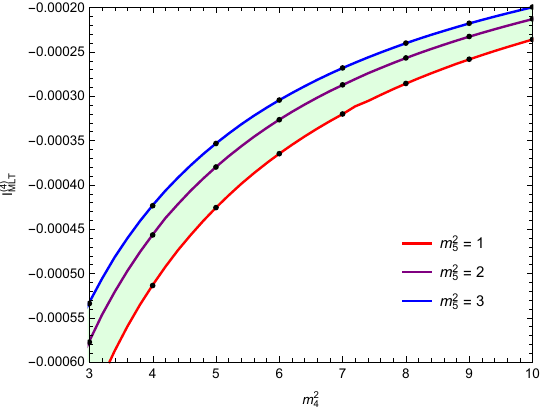}
\includegraphics[scale=0.8]{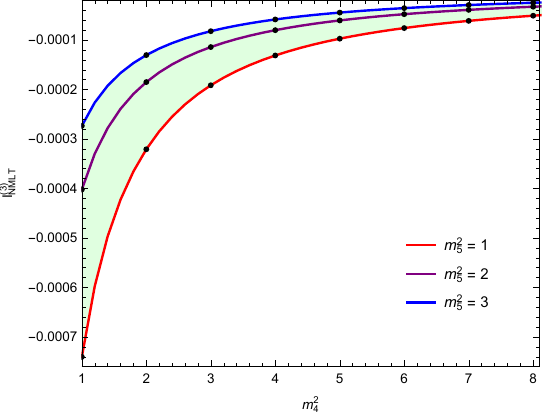}
\includegraphics[scale=0.8]{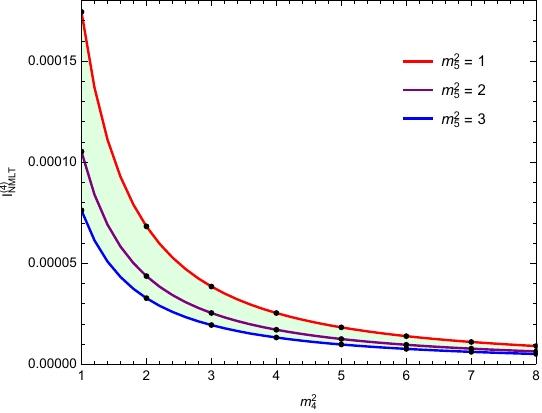}
\includegraphics[scale=0.8]{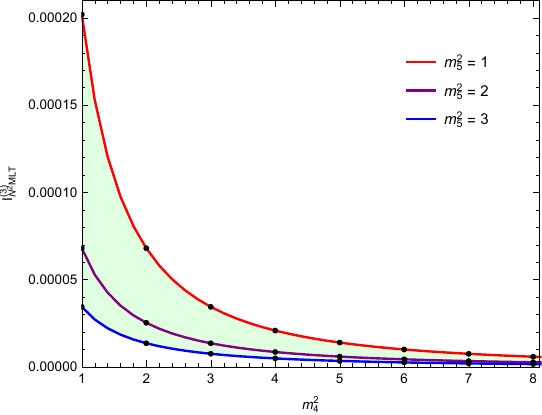}
\includegraphics[scale=0.8]{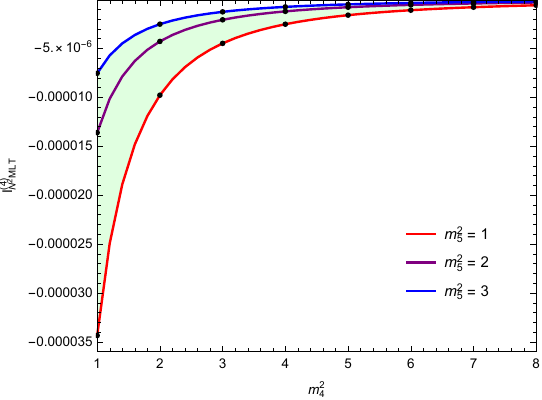}
\caption{Two-dimensional MLT, NMLT and N$^2$MLT at three and four loops, as a function of the internal masses 
$m_4^2$ and $m_5^2$. Solid lines correspond to the analytic results of LTD and dots to the numerical results of \Fiesta.}
\label{fig:2dintegrals}
\end{figure}

We start with the first non-trivial numerical application at $d=2$ space-time dimensions,
in which we perform $L$-loop integrations, one integration per loop.
In order to perform these integrations, 
we embed the integration domain, $\mathbb{R}^{L}$, in a $L$-dimensional sphere. 
This set of integration variables has the features that only one variable goes to infinity,
\begin{align}
\ensuremath{r\in[0,\infty)\,,\,\theta_{1}\in[0,\pi]}\,,\hdots,\,\theta_{L-2}\in[0,\pi]\,,\,\theta_{L-1}\in[0,2\pi]\,,
\end{align}
which we can compactify. Then, its domain is mapped onto $[0,1]$
through to the change of variable, 
\begin{align}
r\to\frac{x}{1-x}\,.
\end{align}
These operations are carried out in \Mathematica~as well as the numerical integration,
which was performed with the built-in function \verb"NIntegrate".

Then, with all the ingredients ready to perform the integrations, 
we evaluate the multi-loop integrals in which the propagators of the lines
$\{1,2,\hdots,L\}$~\eqref{eq:setMLT} have mass $m_4^2$, while the remaining 
ones have mass $m_5^2$. 
Likewise, to test the smooth behaviour of these integrations, 
we scan over $m_4^2\in [1,10]$.
Here and in the following, all kinematic invariants are implicitly given in GeV$^2$.
We then integrate numerically up to four loops the MLT, NMLT and N$^2$MLT topologies
presented in Sec.~\ref{sec:analytic}.
Nevertheless, the extension to higher loops does not originate any 
obstacle within our approach. 

The evaluation of the two-dimensional integrals is shown in Fig.~\ref{fig:2dintegrals},
where solid lines corresponds to the evaluation within LTD, and the 
dots represent the evaluations performed by \Fiesta~and \SecDec.
The evaluation time per point was $\mathcal{O}\left(1''\right)$
in a desktop machine with an Intel i7 (3.4GHz) processor with 8 cores and 16 GB of RAM. 
Additionally, we report that when including more propagators or, equivalently, 
inserting external momenta, the number of integrations for the softwares based on
sector decomposition increases w.r.t. the number of Feynman parameters, instead, 
within LTD, one always has to perform $L$ integrations.

%%%%%%%%%%%%%%%%%%%%%%%%%%%%%%%%%%%%
\subsection{Three- and four-dimensional integrals}
\label{sec:34d-int}

\begin{figure}[thb!]
\centering
\includegraphics[scale=0.8]{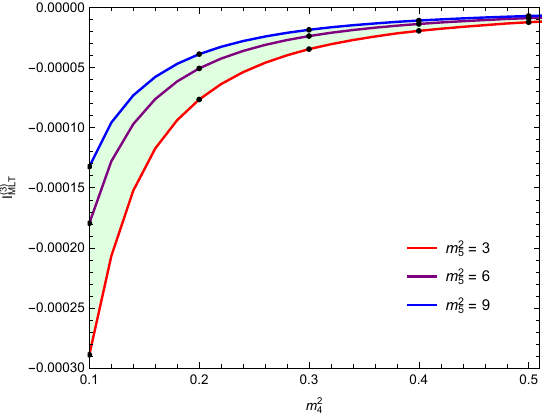}
\includegraphics[scale=0.8]{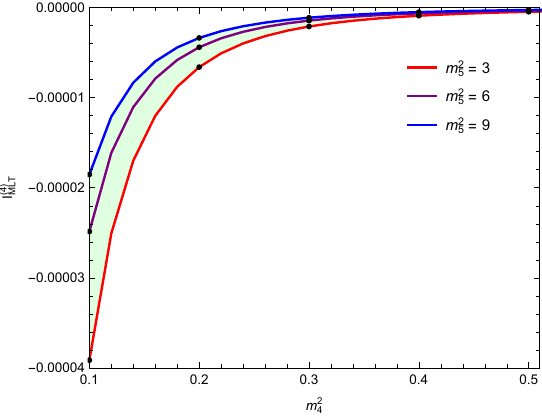}
\includegraphics[scale=0.8]{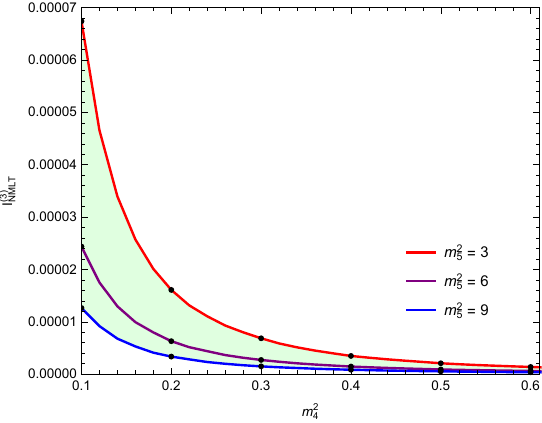}
\includegraphics[scale=0.8]{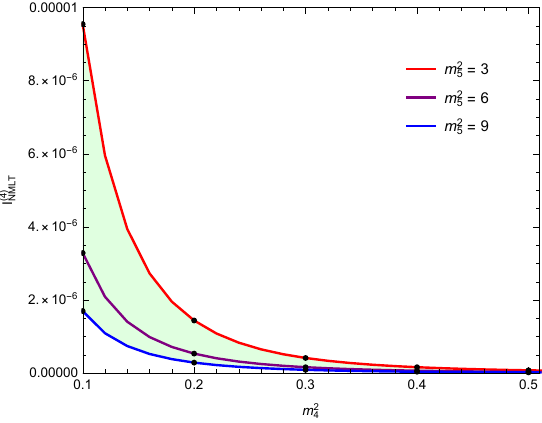}
\includegraphics[scale=0.8]{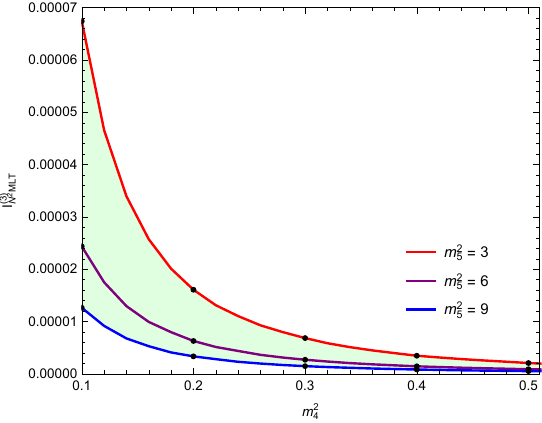}
\includegraphics[scale=0.8]{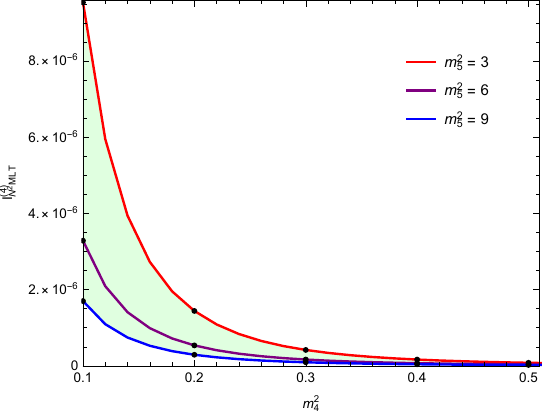}
\caption{Three-dimensional MLT, NMLT and N$^2$MLT at three and four loops, as a function of the internal masses $m_4^2$ and $m_5^2$. 
Solid lines correspond to the analytic results of LTD and dots to the numerical results of \Fiesta.}
\label{fig:3dintegrals}
\end{figure}

\begin{figure}[thb!]
\centering
\includegraphics[scale=0.8]{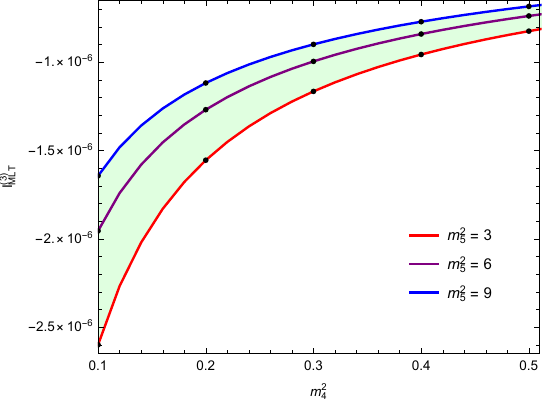}
\includegraphics[scale=0.8]{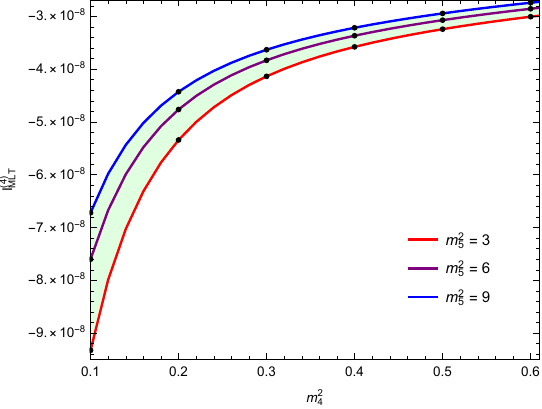}
\includegraphics[scale=0.8]{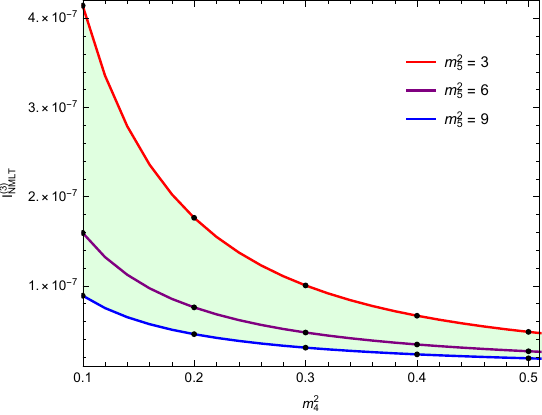}
\includegraphics[scale=0.8]{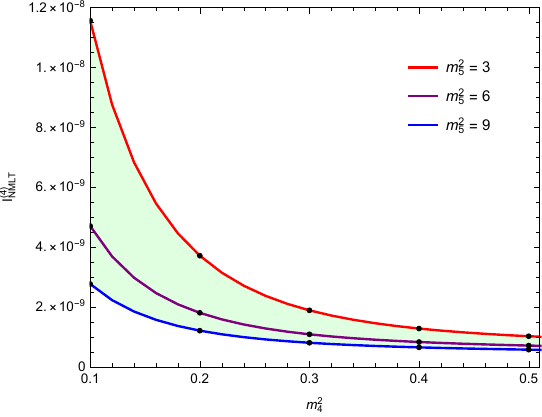}
\includegraphics[scale=0.8]{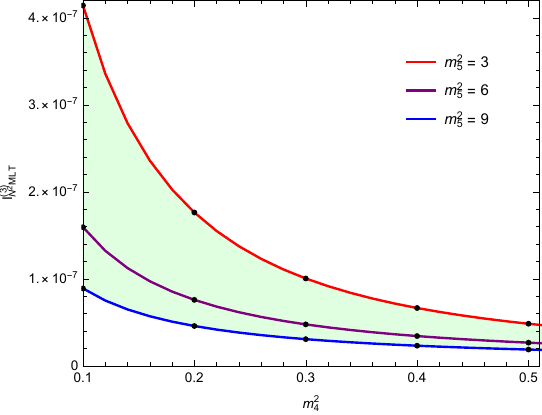}
\includegraphics[scale=0.8]{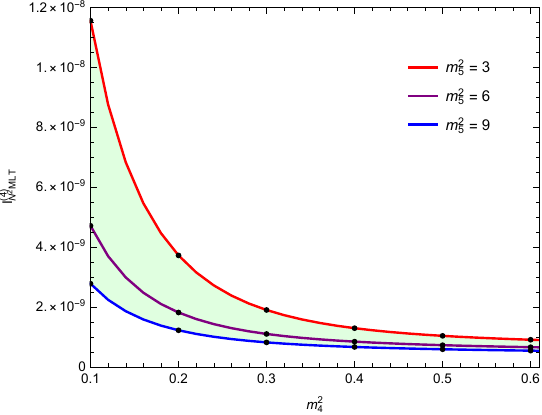}
\caption{Four-dimensional MLT, NMLT and N$^2$MLT at three and four loops, as a function of the internal masses $m_4^2$ and $m_5^2$. 
Solid lines correspond to the analytic results of LTD and dots to the numerical results of \Fiesta.}
\label{fig:4dintegrals}
\end{figure}

It is clear that the number of integrations depends on the dimensionality of the loop momenta.
Hence, we can still use the same procedure of Sec.~\ref{sec:2d-int} to 
express all loop components in terms of spherical coordinates. 
Thanks to the LTD theorem, we pass from Minkowskian to Euclidean space, 
which, in practice, corresponds to work in $\mathbb{R}^{\left(d-1\right)L}$. 
Hence, the embedding in a $\left(d-1\right)L$-dimensional sphere 
can be carried out analogously as in the  former section. 

Alternatively, an equivalent approach consists in treating each loop momentum
independently when doing the change of variables. 
For instance, the integration domain can be separately expressed as follows,
\begin{align}
\mathbb{R}^{\left(d-1\right)L}=\prod_{i=1}^{L}\mathbb{R}^{\left(d-1\right)}\,,
\label{eq:prodsphe}
\end{align}
where each term in the product is the $\left(d-1\right)$-dimensional space of each 
loop momentum. 
The main difference between this approach and the former one relies on how 
the integrand behaves at infinity. In particular, embedding the integrand in a $(d-1)L$-dimensional
sphere allows us to reach and understand this behaviour with a single variable.
Instead, in the product of $(d-1)$-dimensional spheres~\eqref{eq:prodsphe}, 
this behaviour is understood with $L$ variables. 
In the present discussion, we, nonetheless, follow both approaches 
as a double check of our results. 

The numerical integrations within LTD, in $d=3$ and $d=4$,
for the MLT, NMLT and N$^2$MLT configurations with higher powers in the propagators, 
obtained from the causal representations of Sec.~\ref{sec:analytic}, 
\begin{align}
 {\cal A}_{\text{N}^{k-1}\text{MLT}}^{\left(L\right)} \big(1^{2},2^{2},\ldots, L^{2}&, L+1, \dots, L+k \big) \nn \\
&\,=\prod_{i=1}^{L}\frac{\partial}{\partial(q_{i,0}^{\left(+\right)})^{2}} {\cal A}_{\text{N}^{k-1}\text{MLT}}^{\left(L\right)}\left(1,2,\hdots,L+1,
\ldots, L+k \right)\,,
\end{align}
are shown in Fig.~\ref{fig:3dintegrals}
and~\ref{fig:4dintegrals}, respectively. 
In the same way as done in Sec.~\ref{sec:2d-int},
we make a scan in $m_4^2$, by fixing $m_5^2$. 

%%%%%%%%%%%%%%%%%%%%%%%%%%%%%%%%%%%%%%%%%%%%%%%%%%%%%%%%%%%%%%
%
%				CONCLUSIONS
%
%%%%%%%%%%%%%%%%%%%%%%%%%%%%%%%%%%%%%%%%%%%%%%%%%%%%%%%%%%%%%%

\section{Conclusions}
\label{sec:conclusions}

In this paper, we have explicitly elaborated on the analytical structure
of the Maximal (MLT), Next-to-Maximal (NMLT) and Next-to-Next-to-Maximal
(N$^{2}$MLT) loop topologies to all orders. 
We noted that any multi-loop scattering amplitude constructed from the MLT, NMLT and N$^{2}$MLT
topologies, within the loop-tree duality formalism, can always be expressed
in terms of causal propagators only. 
The causal representation contains products of causal propagators  
that can be interpreted as entangled thresholds. 

In order to understand the pattern of the multi-loop MLT, NMLT and N$^{2}$MLT topologies,
we reconstructed their compact analytic expressions
and, thus, elucidated their causal structure.  We made use
of reconstruction of analytic expressions from numerical evaluations
over finite fields, and generated
compact formulae at arbitrary numbers of loops. 
The calculation of the compact causal formulae was explicitly carried out for 
multi-loop vacuum integrals as well as for their extension with external momenta.

We also studied the behaviour of topologies with higher powers in
the propagators. In particular, we noted that the results generated
for topologies with single power propagators give the relevant causal information. 
Hence, due to the explicit dependence in the on-shell loop energies $q_{i,0}^{\left(+\right)}$
of the LTD representation, we defined an operator that raises 
the powers of  propagators starting from the original single-power causal representation.

In view of the compact and simple expressions of MLT, NMLT and N$^{2}$MLT,
we integrated them numerically up to four loops in integer space-time dimensions,
$d=2,3,4$. Since we had to perform $(d-1)L$ integrations, 
we followed two approaches to evaluate these integrals. The first one embeds the
integration domain $\mathbb{R}^{\left(d-1\right)L}$ in $\left(d-1\right)L$ sphere,
whereas the second one embeds the former in $L$ products of 
$\left(d-1\right)$ spheres. We found agreement in both
approaches. Furthermore, we checked our expressions with available softwares
based on sector decomposition, \Fiesta~and \SecDec. 

The algorithm presented in this paper and the interpretation in terms of entangled causal thresholds 
can be extended to other loop integrals with more complex 
internal configurations and to new N$^{k-1}$MLT topologies, with $k>3$, that will appear beyond three loops. 
We recall that a causal representation is expected, since in the end N$^{k-1}$MLT admits a decomposition 
in terms of MLT subtopologies.

%%%%%%%%%%%%%%%%%%%%%%%%%%%%%%%%%%%%%%%%%%%%%%%%%%%%%%%%%%%%%%
%
%				ACKNOWLEDGEMENTS
%
%%%%%%%%%%%%%%%%%%%%%%%%%%%%%%%%%%%%%%%%%%%%%%%%%%%%%%%%%%%%%%

\section*{Acknowledgements}

This work is supported by the Spanish Government (Agencia Estatal de Investigaci\'on) 
and ERDF funds from European Commission (Grant No. FPA2017-84445-P), 
Generalitat Valenciana (Grant No. PROMETEO/2017/053) and the COST Action CA16201 PARTICLEFACE. 
R.J.H.-P. acknowledges support from Departament de F\'isica Te\`orica, Universitat de Val\`encia, CONACyT through
the Project No. A1-S-33202 (Ciencia B\'asica) and Sistema Nacional de Investigadores;
W.J.T.  from Juan de la Cierva program (FJCI-2017-32128), 
and J.J.A.V. from Generalitat Valenciana (GRISOLIAP/2018/101).

\bibliographystyle{JHEP}
\bibliography{refs}

\end{document}